\documentclass{article}
\usepackage[preprint,nonatbib]{neurips}
\usepackage{wrapfig}
\usepackage[inline, shortlabels]{enumitem}
\usepackage[utf8]{inputenc} % allow utf-8 input
\usepackage[T1]{fontenc}    % use 8-bit T1 fonts
\usepackage{hyperref}       % hyperlinks
\usepackage{url}            % simple URL typesetting
\usepackage{booktabs}       % professional-quality tables
\usepackage{amsfonts}       % blackboard math symbols
\usepackage{nicefrac}       % compact symbols for 1/2, etc.
\usepackage{microtype}      % microtypography
\usepackage{xcolor}         % colors
\usepackage{enumitem}[nolistsep]
\usepackage{multirow} 
\usepackage{array}
\usepackage[para]{footmisc}
\usepackage{bbm}
\usepackage{adjustbox}
\usepackage{pgfplots}
\usepackage{graphicx}
\usepackage{caption}
\usepackage{subcaption}
\usepackage{pgfkeys}
\usepackage[normalem]{ulem}
\usepackage{algpseudocode}
\usepackage{algorithm}
\usepackage{amsmath}

\def\addlegendimage{\csname pgfplots@addlegendimage\endcsname}

\AtBeginDocument{%
  \providecommand\BibTeX{{%
    \normalfont B\kern-0.5em{\scshape i\kern-0.25em b}\kern-0.8em\TeX}}}

\newlist{inlinelist}{enumerate*}{1}
\setlist*[inlinelist,1]{label=\roman*),itemjoin={{, }},itemjoin*={{, and }}}

\begin{document}

%\clearheadinfo
\title{SPLADE-v3: New baselines for SPLADE}

\author{%
Carlos Lassance \\
  Cohere (\emph{Work done while at Naver})\\
  cadurosar at gmail dot com \\
  \And
  Hervé Déjean, Thibault Formal, Stéphane Clinchant \\
  Naver Labs Europe \\
  first.lastname at naverlabs dot com
  % examples of more authors
%  \And
%  Nicola Tonellotto \\
%  University of Pisa
  % Address \\
  % \texttt{email} \\
  % \AND
  % Coauthor \\
  % Affiliation \\
  % Address \\
  % \texttt{email} \\
  % \And
  % Coauthor \\
  % Affiliation \\
  % Address \\
  % \texttt{email} \\
  % \And
  % Coauthor \\
  % Affiliation \\
  % Address \\
  % \texttt{email} \\
}

%\institute{Naver Labs Europe, Meylan, France \\ \email{\{carlos.lassance,stephane.clinchant\}@naverlabs.com}}

%


\maketitle

\begin{abstract}

A companion to the release of the latest version of the SPLADE library. We describe changes to the training structure and present our latest series of models -- SPLADE-v3. We compare this new version to BM25, SPLADE++, as well as re-rankers, and showcase its effectiveness via a meta-analysis over more than 40 query sets. SPLADE-v3 further pushes the limit of SPLADE models: it is statistically significantly more effective than both BM25 and SPLADE++, while comparing well to cross-encoder re-rankers. Specifically, it gets more than 40 MRR@10 on the MS MARCO dev set, and improves by $\uparrow 2\%$ the out-of-domain results on the BEIR benchmark.  

\end{abstract}

\section{Introduction}

This technical report is a companion to the release of the latest version of the SPLADE library\footnote{\url{https://github.com/naver/splade}}. Given the improvements stemming from simple modifications to the overall training structure, we believe that it is worth releasing new models -- despite the lack of novelty required for a proper publication. We thus aim to document this new series of models -- named SPLADE-v3 -- and provide the community with better SPLADE ``baselines''. We have been using this new version of the code for most of our recent works.

\section{Better Training}

We detail in the following several improvements that have been made to the training of SPLADE models.

\subsection{Multiple Negatives Per Batch}

Following Tevatron~\cite{gao2022tevatron}, the library now allows training with more than one hard negative. We find that increasing the number of negatives improves the results, especially in the in-domain~\cite{dejean2023benchmarking} setting, but does not add much to out-of-domain generalization. We use negatives coming from a SPLADE++~\cite{formal2022distillation} model, and consider 100 negatives -- 50 from the top-50 and 50 chosen at random from the top-$1k$.

\subsection{Better Distillation Scores}

To further improve SPLADE's effectiveness, we use an ensemble of cross-encoder re-rankers to generate our distillation scores -- instead of the standard approach relying on a single  model~\cite{lassance2022efficiency,formal2022distillation,formal2021splade}\footnote{Especially, \texttt{cross-encoder/ms-marco-MiniLM-L-6-v2}}. We generate two types of scores: \begin{enumerate*}
    \item the simple ensemble of scores, and
    \item the ``rescored'' version, where we use affine transformations to make some of the data statistics (average and std score values) similar to the ones encountered in the previous distillation setting~\footnote{\url{https://huggingface.co/datasets/sentence-transformers/msmarco-hard-negatives} }
\end{enumerate*}.
We use the following open-source models on HuggingFace to generate the scores: 
\begin{enumerate}
    \item \texttt{cross-encoder/ms-marco-MiniLM-L-6-v2} $\clubsuit$
    \item \texttt{naver/trecdl22-crossencoder-rankT53b-repro}
    \item \texttt{naver/trecdl22-crossencoder-debertav3}
    \item \texttt{naver/trecdl22-crossencoder-debertav2}
    \item \texttt{naver/trecdl22-crossencoder-electra}
\end{enumerate}

Where the first one ($\clubsuit$) is the one that generated the scores for SPLADE++~\cite{formal2022distillation}, and the remaining ones are models we trained on MS MARCO for the 2022 edition of the TREC Deep Learning task \cite{Craswell2022OverviewOT}.

We feed each of the $500k$ queries of the training set of MS MARCO -- paired with each of the 100 negatives and the positive(s) -- to the re-rankers. The scores are then normalized per query using the min-max aggregation from ranx~\cite{bassani2022ranx}. This generates our ``ensemble'' scores. To generate our ``rescored'' scores, we look into the statistics of the ensemble scores and use an affine transformation so that the average score and the standard deviation closely mimic the previous scores ($\clubsuit$). We notice empirically that changing the distribution helps when using distillation -- especially in the case of MarginMSE \cite{hofstätter2021improving} -- but we didn't investigate further into why this happens.

\subsection{Two Distillation Losses}

In the context of IR, two main distillation losses have proven to be effective: KL-Div~\cite{lin2020distilling} (used for Eff-SPLADE~\cite{lassance2022efficiency}) and MarginMSE~\cite{hofstätter2021improving} (used for SPLADE v2~\cite{formal2021splade} and SPLADE++~\cite{formal2021splade}). Given the extra negatives, we noticed \emph{empirically} that the MarginMSE (resp. KL-Div) focused more on Recall (resp. Precision). We then chose to combine both, with different weights ($\lambda_{KL}=1$ for KL-Div, $\lambda_{MSE}=0.05$ for MarginMSE -- based on cross-validation), which overall led to better results. 

% -- because one has many more distances between negatives than a positive and a negative --

\subsection{Further Fine-Tuning SPLADE}

We also noticed that starting from SPLADE++\texttt{SelfDistil}\footnote{\url{naver/splade-cocondenser-selfdistil}} -- which exhibits slight zero-shot boosts when compared to SPLADE++\texttt{EnsembleDistil} ~\cite{formal2022distillation} -- and applying the previous changes led to better effectiveness when compared to starting from a CoCondenser~\cite{gao-callan-2022-unsupervised} or a DistilBERT\cite{Victor2019} checkpoint. We are still not sure about the cause(s) of this effect, but we believe that a sort of curriculum learning -- as the one investigated in {\it Zeng et al.}~\cite{zeng2022curriculum} -- could happen and lead to the observed improvements, but it still needs to be better investigated.

\section{A New Baseline, SPLADE-v3}

\paragraph{SPLADE-v3} The base SPLADE-v3 model\footnote{\texttt{naver/splade-v3}} starts from SPLADE++\texttt{SelfDistil}, and is trained with a mix of KL-Div and MarginMSE, with 8 negatives per query sampled from SPLADE++\texttt{SelfDistil}. All the other hyperparameters are similar to previous SPLADE iterations. Importantly, note that in all of our experiments, we use the original MS MARCO collection \emph{without the titles} \cite{lassance2023tale_arxiv,lassance2023tale_sigir}.

\paragraph{Evaluation} To assess the effectiveness of the model, we use the meta-analysis procedure introduced in RANGER~\cite{sertkan2023ranger,sertkan2023exploring}. We use up to 44 query sets -- relying on the \texttt{ir\_datasets} library~\cite{macavaney:sigir2021-irds} -- coming from different datasets, including \begin{enumerate*}
    \item MS MARCO passages (4 query sets),
    \item MS MARCO v2 passages (4 query sets),
    \item BEIR (13 query sets),
    \item LoTTE (12 query sets),
    \item Antique,
    \item TREC-CAR (y1) (2 query sets),
    \item Natural Questions,
    \item TriviaQA,
    \item TREC-TB (3 query sets), and
    \item TREC-MQ (2 query sets).
    \end{enumerate*} We use nDCG*@10 to measure effectiveness, where nDCG* stands for the nDCG considering only the judged documents (encountered in the retrieved top-$k$) if the dataset has both positive and negative judgments -- otherwise, we use the standard nDCG@10.

\paragraph{Comparison to BM25}

\begin{figure}
    \centering
    \adjustbox{max width=0.6\textwidth}{%
        \includegraphics{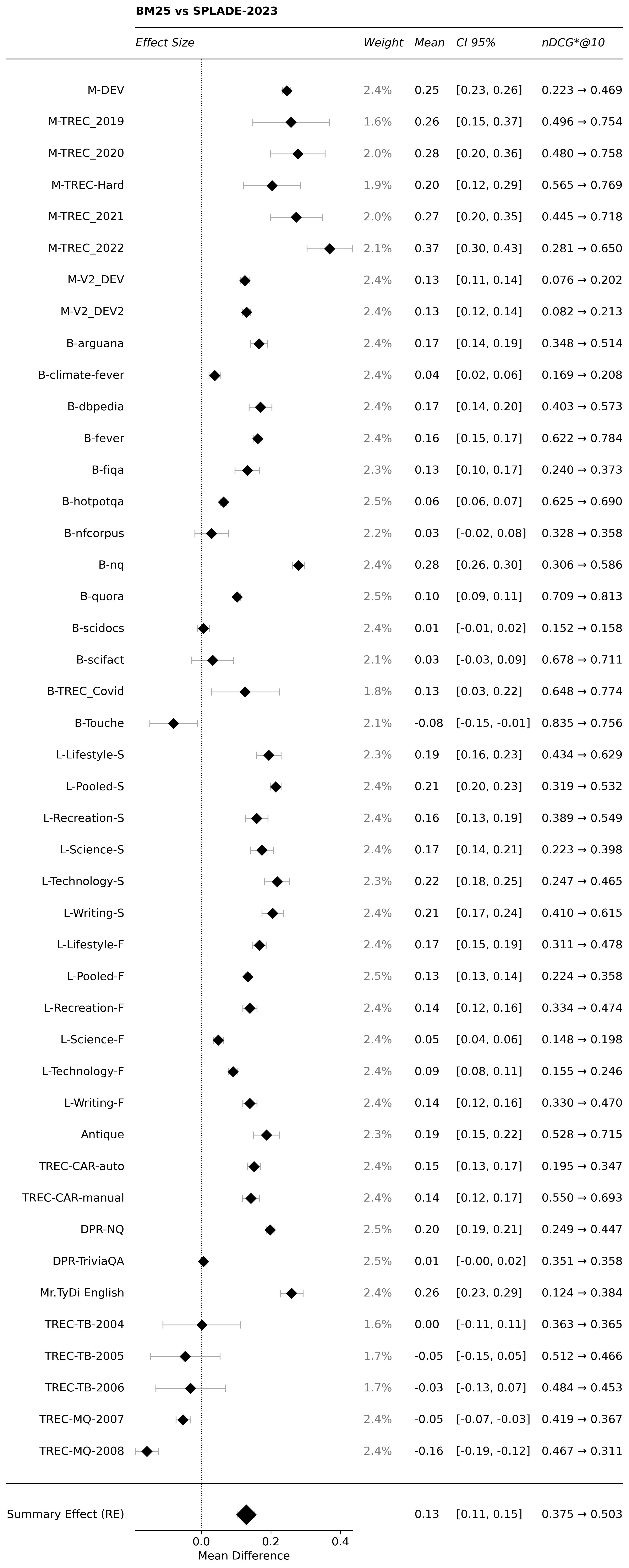}
        }
        \caption{Meta-analysis comparison of SPLADE-v3 and BM25. }
    \label{fig:bm25}
\end{figure}

First, we compare our method to BM25 and present the resulting meta-analysis in Figure~\ref{fig:bm25}. We notice statistically significant improvements in most of the 44 query sets, with only 3 query sets presenting a statistical decrease in effectiveness. These query sets are Webis Touché-2020 and the two TREC-MQ query sets. For Touché-2020, we are still unsure what is the actual issue, but this observation is recurrent with learned ranking models~\cite{formal2022distillation,lassance2022efficiency,thakur2021beir}. For TREC-MQ, there could be an issue with the long documents that may need to be decomposed into passages. Notice the large summary effect, meaning that over the whole set of comparisons, SPLADE-v3 vastly outperforms BM25 (even if it is less efficient).

\paragraph{Comparison to SPLADE++\texttt{SelfDistil}}

\begin{figure}
    \centering
    \adjustbox{max width=0.6\textwidth}{%
        \includegraphics{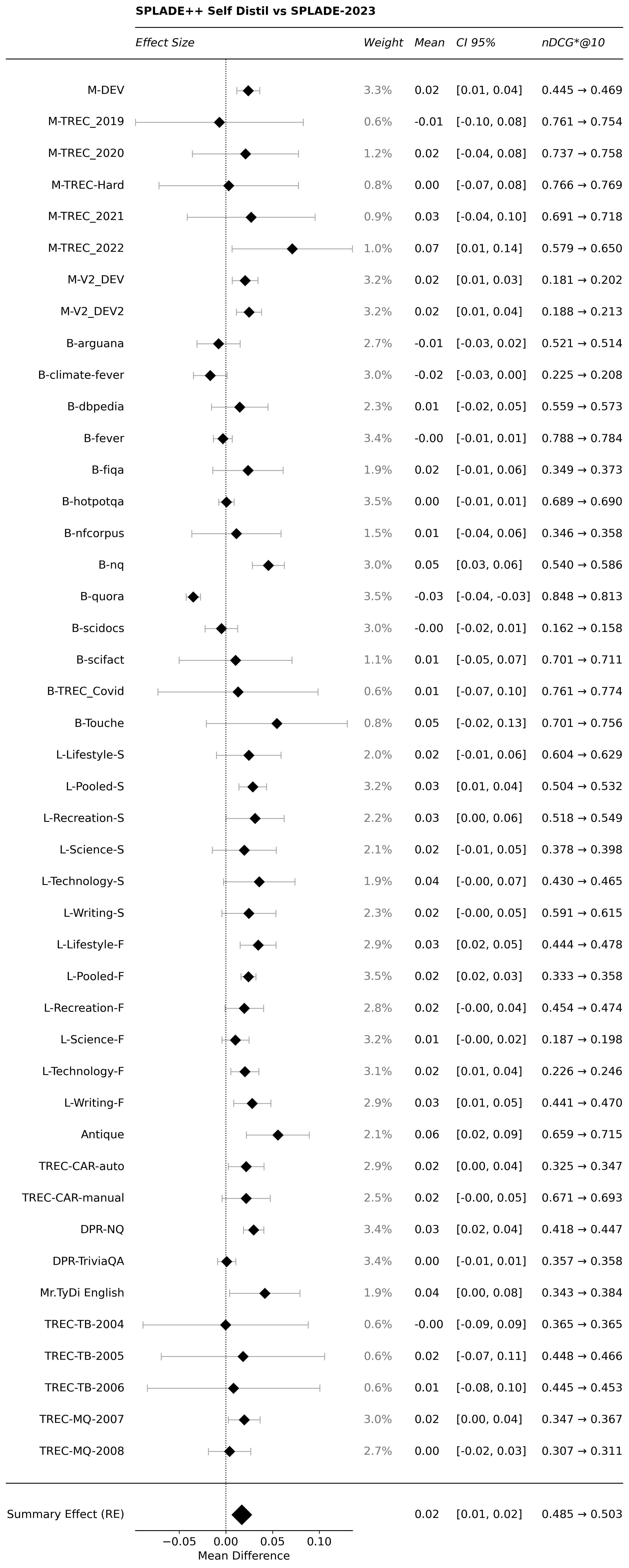}
        }
        \caption{Meta-analysis comparison of SPLADE-v3 and SPLADE++\texttt{SelfDistil}. }
    \label{fig:splades}
\end{figure}

We now compare SPLADE-v3 to the previous SPLADE model used at initialization -- SPLADE++\texttt{SelfDistil}. Ideally, there should not be any loss in effectiveness for any of the tested query sets. We present the meta-analysis in Figure~\ref{fig:splades}. We notice that only Quora suffered from a significant decrease in effectiveness, with most other datasets presenting a gain of effectiveness, with the overall summary effect being positive towards the new baseline.

\paragraph{Comparison to re-rankers}

\begin{figure}
    \centering
    \adjustbox{max width=0.6\textwidth}{%
        \includegraphics{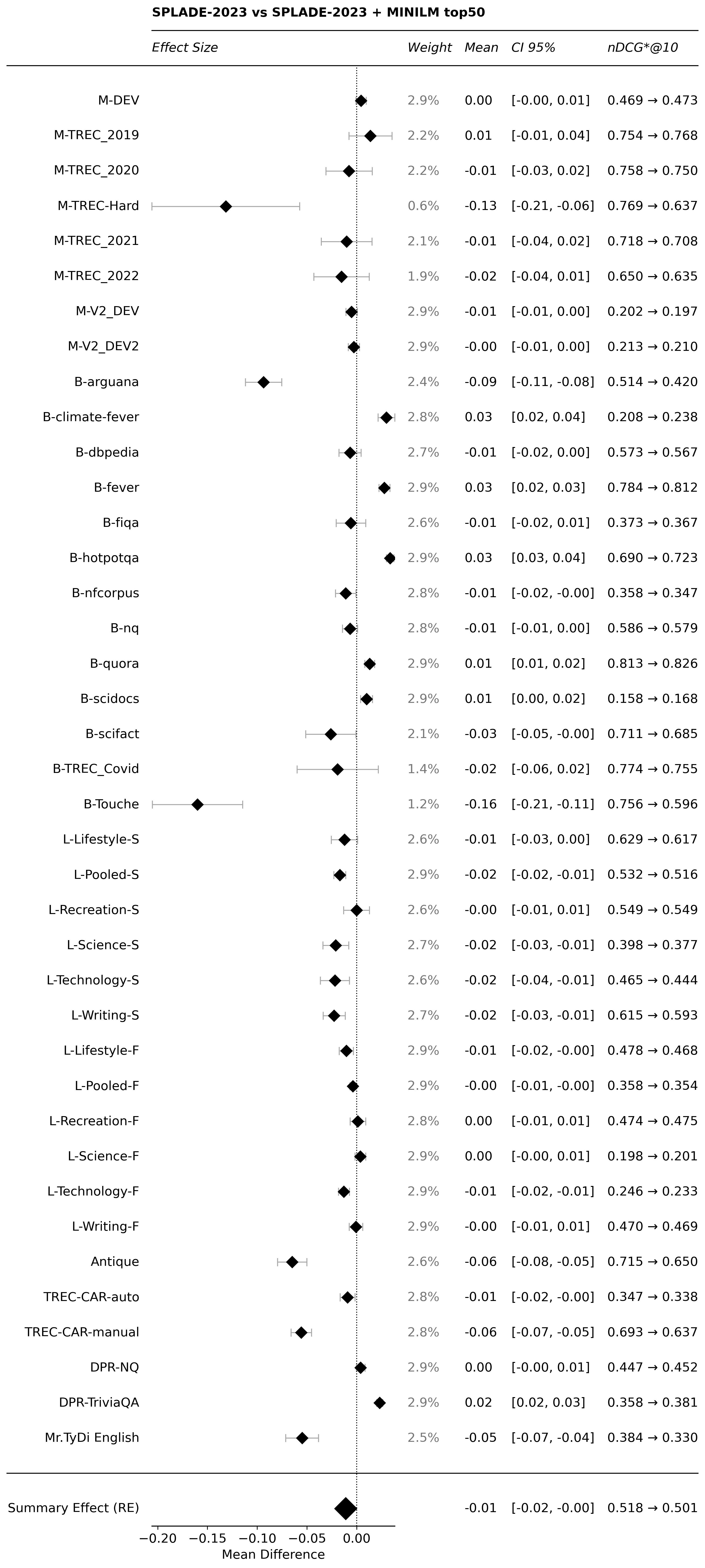}
        }
        \caption{Meta-analysis comparison of SPLADE-v3 and MiniLM (re-ranking the top-50 returned by SPLADE-v3). }
    \label{fig:minilm}
\end{figure}

\begin{figure}
    \centering
    \adjustbox{max width=0.6\textwidth}{%
        \includegraphics{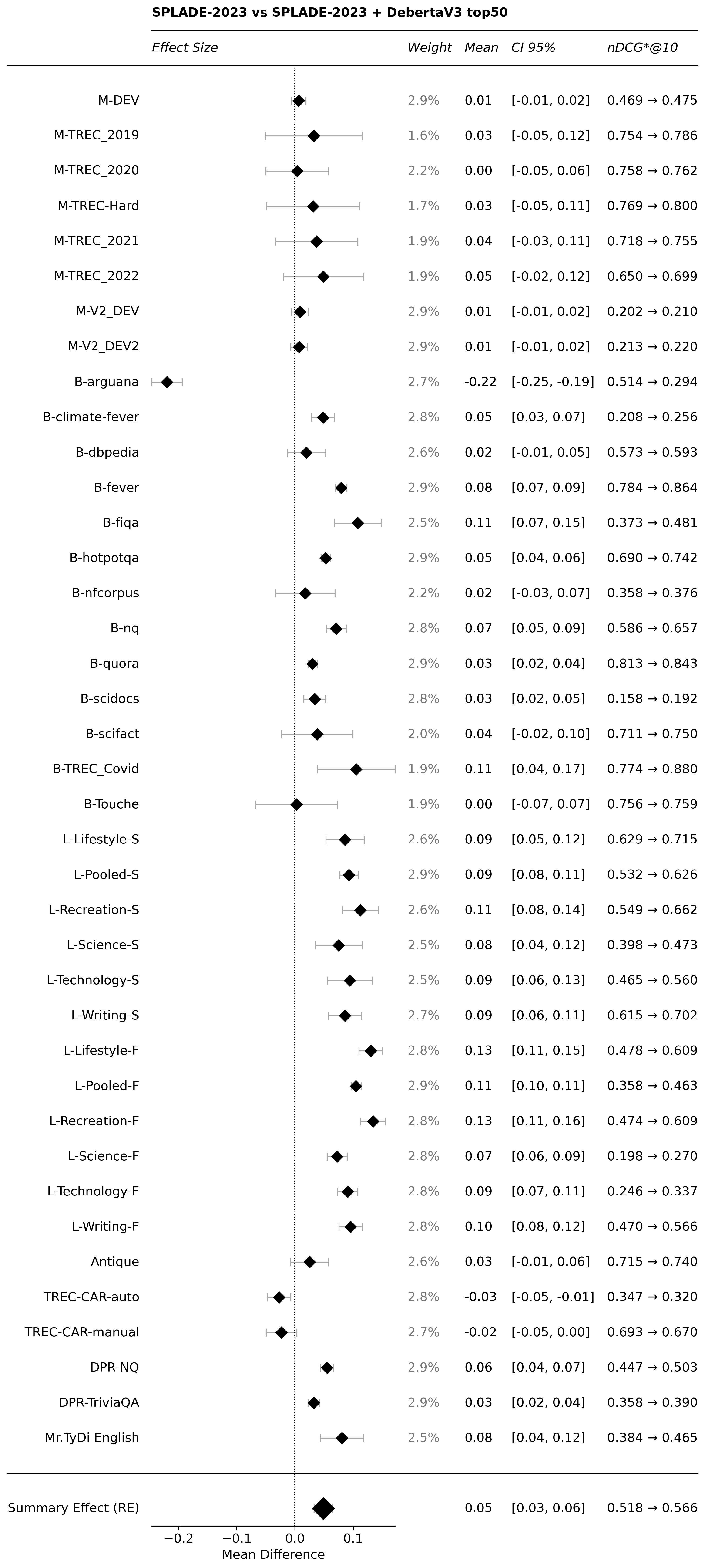}
        }
        \caption{Meta-analysis comparison of SPLADE-v3 and DeBERTaV3 (re-ranking the top-50 returned by SPLADE-v3).}
    \label{fig:deberta}
\end{figure}

We finally compare SPLADE-v3 to cross-encoder re-rankers. More specifically, we consider two models that re-rank the top $k=50$ documents returned by SPLADE-v3: MiniLM\footnote{\texttt{cross-encoder/ms-marco-MiniLM-L-6-v2}} and DeBERTaV3\footnote{\texttt{naver/trecdl22-crossencoder-debertav3}} -- we present the results in Figure~\ref{fig:minilm} and Figure~\ref{fig:deberta} respectively. Note that higher $k$ could be used for re-ranking -- but we believe that re-ranking 50 documents already constitutes a good efficiency-effectiveness trade-off, especially when re-ranking a well-tuned first-stage retriever. For MiniLM, we notice that the summary effect is close to 0 when we consider a 95~\% confidence interval and that there is not much difference between the original results and the re-ranked ones -- except for a few datasets that could just be ``outliers'' in the effectiveness of MiniLM. However, in the case of DeBERTaV3, we see the opposite: for most query sets the re-ranker is able to outperform SPLADE-v3 -- except for ArguAna whose ``counter-argument'' task might be more intricate for a re-ranker. 

\section{SPLADE-v3-DistilBERT, SPLADE-v3-Lexical and SPLADE-v3-Doc}

In addition, we also release three other SPLADE-v3 variants:
\begin{enumerate}
    \item SPLADE-v3-DistilBERT\footnote{\texttt{naver/splade-v3-distilbert}}, which instead starts training from DistilBERT -- and thus has a smaller inference ``footprint''.
    
    \item SPLADE-v3-Lexical\footnote{\texttt{naver/splade-v3-lexical}}, for which we remove query \emph{expansion}, thus reducing the retrieval FLOPS (and improving efficiency)~\cite{10.1145/3634912}. 
    \item SPLADE-v3-Doc\footnote{\texttt{naver/splade-v3-doc}}, which starts training from CoCondenser, and where no computation is done for the query -- which can be seen as a simple binary Bag-of-Words~\cite{formal2021splade,10.1145/3634912}. 
    \end{enumerate}
%The training configs are available in ``conf/main\_config/hf*''.

Table~\ref{tab:baselines} summarizes the results as averages over datasets -- detailed results over the set of 13 BEIR datasets can be found in Table~\ref{tab:beir}. We note that SPLADE-v3-Lexical is (very) effective on MS MARCO as well as LoTTE, but struggles on BEIR (out-of-domain). While the DistilBERT version is a clear downgrade from the BERT version, it remains however more effective than the lexical version on BEIR. SPLADE-v3-Doc is the less effective approach overall, especially in ``zero-shot'', showing that (even) a minimal amount of computation on the query side is important. However, its performance remains quite competitive w.r.t. state-of-the-art dense bi-encoders, especially given its efficiency (no query encoding, and a short number of posting lists to traverse).

\begin{table}[ht]
\caption{Comparison of results as averages per dataset. We report MRR@10 for MS MARCO (MSM), nDCG@10 for TREC, mean nDCG@10 for BEIR (13 datasets), and mean Success@5 over all non-pooled subsets of the Forum (LoTTE-F) and Search (LoTTE-S) tasks for LoTTE \cite{santhanam-etal-2022-colbertv2}. We also report the FLOPS measure as a loose indicator of efficiency~\cite{splade}.}
\label{tab:baselines}
\adjustbox{max width=1.0\textwidth}{%
\begin{tabular}{lccccccc}
\toprule 
Model                                     & MSM & TREC19 & TREC20 & BEIR 13 & LoTTE-S & LoTTE-F & FLOPS \\ \midrule 
SPLADE++SD  & 37.6    & 73.0  & 71.8  & 50.7   & -        & -       & 1.4   \\ \midrule
SPLADE-v3  & 40.2    & 72.3  & 75.4  & 51.7   & 74.7        & 66.0       & 1.2   \\
SPLADE-v3-DistilBERT & 38.7    & 75.2  & 74.4  & 50.0   & 70.3        & 62.8       & 1.4   \\
SPLADE-v3-Lexical & 40.0    & 71.2  & 73.6  & 49.1   & 74.2        & 64.5       & 0.6   \\ 
SPLADE-v3-Doc    & 37.8      & 71.5  &   70.3 &  47.0 &71.1  & 59.0& 1.4\\ 
\bottomrule
\end{tabular}
}
\end{table}

\begin{table}[ht]
\caption{nDCG@10 over the set of 13 datasets of BEIR~\cite{thakur2021beir}.}
\label{tab:beir}
\adjustbox{max width=1.0\textwidth}{%
\begin{tabular}{lccccc}
\toprule
Dataset        & SPLADE++SD & SPLADE-v3 & SPLADE-v3-DistilBERT & SPLADE-v3-Lexical & SPLADE-v3-Doc\\ \midrule
ArguAna        & 51.8 & 50.9        & 48.4                   & 52.7         & 46.7    \\
Climate-FEVER  & 23.7 & 23.3        & 22.8                   & 21.8         & 15.9    \\
DBPedia-entity & 43.6 & 45.0        & 42.6                   & 42.8         & 36.1     \\
FEVER          & 79.6 & 79.6        & 79.6                   & 78.5         & 68.9     \\
FiQA-2018           & 34.9 & 37.4        & 33.9                   & 36.4         & 33.6      \\
HotpotQA       & 69.3 & 69.2        & 67.8                   & 68.5         & 66.9    \\
NFCorpus       & 34.5 & 35.7        & 34.8                   & 34.7         & 33.8     \\
NQ             & 53.3 & 58.6        & 54.9                   & 56.1         & 52.1      \\
Quora          & 84.9 & 81.4        & 81.7                   & 73.4         & 77.5     \\
SCIDOCS        & 16.1 & 15.8        & 14.8                   & 15.9         & 15.2     \\
SciFact        & 71.0 & 71.0        & 68.5                   & 71.5         & 68.8      \\
TREC-COVID     & 72.5 & 74.8        & 70.0                   & 63.6         & 68.1      \\
Touché-2020    & 24.2 & 29.3        & 30.1                   & 22.7         &27.0       \\ \midrule
Average        & 50.7 & 51.7        & 50.0                   & 49.1   & 47.0\\ \bottomrule            
\end{tabular}
}
\end{table}

\section{Conclusion}

This technical report describes the release of SPLADE-v3 models. We have shown through extensive evaluations that this new series of SPLADE models is statistically significantly more effective than previous iterations. In most query sets -- including zero-shot settings -- SPLADE-v3 outperforms BM25 and can even rival some re-rankers.

%\balance
\bibliographystyle{abbrv}
\bibliography{sample-base}

\begin{thebibliography}{10}

\bibitem{bassani2022ranx}
E.~Bassani.
\newblock ranx: A blazing-fast python library for ranking evaluation and comparison.
\newblock In {\em European Conference on Information Retrieval}, pages 259--264. Springer, 2022.

\bibitem{Craswell2022OverviewOT}
N.~Craswell, B.~Mitra, E.~Yilmaz, D.~F. Campos, J.~Lin, E.~M. Voorhees, and I.~Soboroff.
\newblock Overview of the trec 2022 deep learning track.
\newblock In {\em Text Retrieval Conference}, 2022.

\bibitem{dejean2023benchmarking}
H.~D{\'e}jean, S.~Clinchant, C.~Lassance, S.~Lupart, and T.~Formal.
\newblock Benchmarking middle-trained language models for neural search.
\newblock {\em arXiv preprint arXiv:2306.02867}, 2023.

\bibitem{formal2021splade}
T.~Formal, C.~Lassance, B.~Piwowarski, and S.~Clinchant.
\newblock Splade v2: Sparse lexical and expansion model for information retrieval, 2021.

\bibitem{formal2022distillation}
T.~Formal, C.~Lassance, B.~Piwowarski, and S.~Clinchant.
\newblock From distillation to hard negative sampling: Making sparse neural ir models more effective.
\newblock In {\em Proceedings of the 45th International ACM SIGIR Conference on Research and Development in Information Retrieval}, pages 2353--2359, 2022.

\bibitem{10.1145/3634912}
T.~Formal, C.~Lassance, B.~Piwowarski, and S.~Clinchant.
\newblock Towards effective and efficient sparse neural information retrieval.
\newblock {\em ACM Trans. Inf. Syst.}, dec 2023.
\newblock Just Accepted.

\bibitem{splade}
T.~Formal, B.~Piwowarski, and S.~Clinchant.
\newblock {SPLADE: Sparse Lexical and Expansion Model for First Stage Ranking}.
\newblock In {\em Proc. SIGIR}, page 2288–2292, 2021.

\bibitem{gao-callan-2022-unsupervised}
L.~Gao and J.~Callan.
\newblock Unsupervised corpus aware language model pre-training for dense passage retrieval.
\newblock In S.~Muresan, P.~Nakov, and A.~Villavicencio, editors, {\em Proceedings of the 60th Annual Meeting of the Association for Computational Linguistics (Volume 1: Long Papers)}, pages 2843--2853, Dublin, Ireland, May 2022. Association for Computational Linguistics.

\bibitem{gao2022tevatron}
L.~Gao, X.~Ma, J.~Lin, and J.~Callan.
\newblock Tevatron: An efficient and flexible toolkit for dense retrieval.
\newblock {\em arXiv preprint arXiv:2203.05765}, 2022.

\bibitem{hofstätter2021improving}
S.~Hofstätter, S.~Althammer, M.~Schröder, M.~Sertkan, and A.~Hanbury.
\newblock Improving efficient neural ranking models with cross-architecture knowledge distillation, 2021.

\bibitem{lassance2022efficiency}
C.~Lassance and S.~Clinchant.
\newblock An efficiency study for splade models.
\newblock In {\em Proceedings of the 45th International ACM SIGIR Conference on Research and Development in Information Retrieval}, pages 2220--2226, 2022.

\bibitem{lassance2023tale_arxiv}
C.~Lassance and S.~Clinchant.
\newblock The tale of two ms marco -- and their unfair comparisons, 2023.

\bibitem{lassance2023tale_sigir}
C.~Lassance and S.~Clinchant.
\newblock The tale of two msmarco - and their unfair comparisons.
\newblock In {\em Proceedings of the 46th International ACM SIGIR Conference on Research and Development in Information Retrieval}, SIGIR '23, page 2431–2435, New York, NY, USA, 2023. Association for Computing Machinery.

\bibitem{lin2020distilling}
S.-C. Lin, J.-H. Yang, and J.~Lin.
\newblock Distilling dense representations for ranking using tightly-coupled teachers, 2020.

\bibitem{macavaney:sigir2021-irds}
S.~MacAvaney, A.~Yates, S.~Feldman, D.~Downey, A.~Cohan, and N.~Goharian.
\newblock Simplified data wrangling with ir\_datasets.
\newblock In {\em SIGIR}, 2021.

\bibitem{Victor2019}
V.~Sanh, L.~Debut, J.~Chaumond, and T.~Wolf.
\newblock Distilbert, a distilled version of bert: smaller, faster, cheaper and lighter, 10 2019.

\bibitem{santhanam-etal-2022-colbertv2}
K.~Santhanam, O.~Khattab, J.~Saad-Falcon, C.~Potts, and M.~Zaharia.
\newblock {C}ol{BERT}v2: Effective and efficient retrieval via lightweight late interaction.
\newblock In M.~Carpuat, M.-C. de~Marneffe, and I.~V. Meza~Ruiz, editors, {\em Proceedings of the 2022 Conference of the North American Chapter of the Association for Computational Linguistics: Human Language Technologies}, pages 3715--3734, Seattle, United States, July 2022. Association for Computational Linguistics.

\bibitem{sertkan2023ranger}
M.~Sertkan, S.~Althammer, and S.~Hofst{\"a}tter.
\newblock Ranger: A toolkit for effect-size based multi-task evaluation.
\newblock {\em arXiv preprint arXiv:2305.15048}, 2023.

\bibitem{sertkan2023exploring}
M.~Sertkan, S.~Althammer, S.~Hofst{\"a}tter, P.~Knees, and J.~Neidhardt.
\newblock Exploring effect-size-based meta-analysis for multi-dataset evaluation.
\newblock 2023.

\bibitem{thakur2021beir}
N.~Thakur, N.~Reimers, A.~R{\"u}ckl{\'e}, A.~Srivastava, and I.~Gurevych.
\newblock Beir: A heterogenous benchmark for zero-shot evaluation of information retrieval models.
\newblock {\em arXiv preprint arXiv:2104.08663}, 2021.

\bibitem{zeng2022curriculum}
H.~Zeng, H.~Zamani, and V.~Vinay.
\newblock Curriculum learning for dense retrieval distillation.
\newblock In {\em Proceedings of the 45th International ACM SIGIR Conference on Research and Development in Information Retrieval}, pages 1979--1983, 2022.

\end{thebibliography}

\end{document}